# Universal Numeric Segmented Display

## Md. Abul Kalam Azad, Rezwana Sharmeen and S. M. Kamruzzaman

Department of Computer Science & Engineering,
International Islamic University Chittagong, Chittagong, Bangladesh.
email {azadarif, r_sharmin_79, smk_iiuc}@yahoo.com

**Abstract**

*Segmentation display plays a vital role to display numerals. But in today's world matrix display is also used in displaying numerals. Because numerals has lots of curve edges which is better supported by matrix display. But as matrix display is costly and complex to implement and also needs more memory, segment display is generally used to display numerals. But as there is yet no proposed compact display architecture to display multiple language numerals at a time, this paper proposes uniform display architecture to display multiple language digits and general mathematical expressions with higher accuracy and simplicity by using a 18-segment display, which is an improvement over the 16 segment display.*

**Keywords**

18-segment display, Latin, Greek, Roman, Arabic, Bengali, Chinese, Hebrew, Devanagari, Tamil, Tibeten, Gujrati, Telegu, Mathematical signs.

## INTRODUCTION

Segment display is now a widely used display method in electronic devices. So far 7-segment display is used for English digits, those are Latin digits and 10-segment display is proposed for Bengali digits [7, 10], 8-segment display is proposed for English and Bengali digits [11], 12-segment display is proposed for Arabic digits [10]. But these segment architectures are different to each other and we cannot use the same architecture for uniform display of multiple languages, that's why we have proposed 18-segment display as a uniform architecture to display multiple numerals at a time and to display numerical expressions. This paper proposes uniform display architecture to display multiple language digits and general mathematical expressions with higher accuracy and simplicity by using 18-segment display, which is an improvement over the 16 segment display. Our proposed 18-segment display can be used to display numerals of twelve languages and mathematical expression. We have used 18-segment display for this purpose which is an improvement over 16-segment display. Our 18-segment display requires two more segments in comparison to the common 16-segment display. As 16-segment display is very common in practice and quite cheap and needs fewer gates to implement that's why we have proposed to use 18-segment display with a bit modification. Moreover by using 18-segment displays we may eliminate the need of a completely new specially fabricated segment display. By adding two segments 16-segment display can be easily converted to 18-segment display for the above reasons we have used 18-segments for displaying multi language digits and mathematical expressions.

We know that, lots of languages use same numeric symbol to represent numerals that's why we categorized the language and proposed a display model which will be able two display the numerals those are widely used in the world. In the case of language selection we also consider the population covered by each language numeral. Our proposed system can be able to display numerals of the following languages:- "Latin, Greek, Roman, Arabic, Bengali, Chinese, Hebrew, Devanagari, Tamil, Tibeten, Gujrati, Telegu". We found that the Latin numerals are used in English, Russian, and French etc. Similarly Arabic numerals are used in Spanish and Urdu etc. Also Devanagari is used in Hindi and Sanskrit etc. We also considered Chinese and Bengali on the perspective of population. Greek, Roman and Hebrew languages are selected on the basis of supremeness, because these three languages are on of the oldest languages that are used in the world now. We also selected Tamil, Tibeten, Gujrati, Telegu on the basis of population, sociocultural and on the basis of educational and life standard.

## PROPOSED SEGMENTED DISPLAY

In the current world 7-segment display is the all in all in Latin type numeral display but 7-segment display or a slight modification of 7-segment display can not be used as an universal numeric display. That's why we decide to use 16-segment display because 16-segment display is also one of the top most popular segmented display. And finally to support multiple language numerals we added some segments with 16-segment display.

### Proposed Segment Architecture

Our proposed 18-segment display is an improvement over the previously designed 16-segment. We have added two extra segments in the 16-segment display. The segments are 1) n segment and 2) p segment. The n segment is placed in the upper part and the segment p is placed in the lower layer part of the display. Finally the segment display takes the form of an 18-segment display. The 18-segment display architecture is shown below, Figure 1 shows the traditional 16-segment display and the Figure 2 shows the proposed 18-segment display:

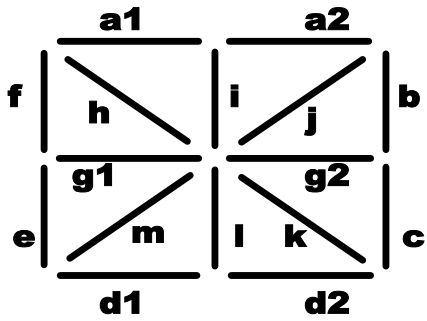

Figure 1: 16-Segment Display

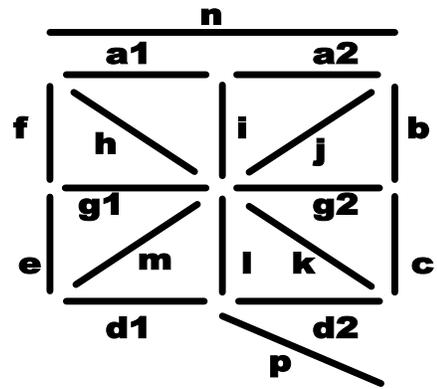

Figure 2: 18-Segment Display

**Segment Patterns Different Language Numerals**
The pattern of different language numeral are shown in the below tables. In the following tables "D Val" stands for Digit Value, "Act Sym" stands for Actual Symbol, "Seg Pat" stands for Segment Pattern, and "Com Vec" stands for Combination Vector.

Table 1: Representation of English Numerals

| D Val | Act Sym | Seg Pat | Com Vec | D Val | Act Sym | Seg Pat | Com Vec |
|---|---|---|---|---|---|---|---|
| 0 | 0 | | { a2,b,c,d2,l,i } | 1 | 1 | | { b,c } |
| 2 | 2 | | { a2,b,g2,l,d2 } | 3 | 3 | | { a2,b,c,g2,d2} |
| 4 | 4 | | { i,g2,b,c } | 5 | 5 | | { a2,i,g2,c,d2 } |
| 6 | 6 | | { a2,i,l,d2,c,g2 } | 7 | 7 | | { a2,b,c } |
| 8 | 8 | | {a2,b,c,d2,l,i,g2 } | 9 | 9 | | {a2,b,c,d2,i,g2} |

Table 2: Representation of Bengali Numerals

| D Val | Act Sym | Seg Pat | Com Vec | D Val | Act Sym | Seg Pat | Com Vec |
|---|---|---|---|---|---|---|---|
| 0 | ০ | | {a1,a2,b,c,d2,d1,e,f } | 1 | ১ | | {a1,a2,b,c,d2,d1,e,g1,l } |
| 2 | ২ | | {a1,a2,b,g2,g1,e,d1,d2 } | 3 | ৩ | | {a2,b,c,d2,d1,e,f,i,g2 } |
| 4 | ৪ | | {a1,a2,b,c,d2,d1,e,f,g1,g2} | 5 | ৫ | | {a1,i,g2,c,d2,d1,e,f } |
| 6 | ৬ | | {i,g2,c,d2,d1,e,f } | 7 | ৭ | | {a1,a2,b,c,g2,g1,f } |
| 8 | ৮ | | { f,e,d1,l,g1,j} | 9 | ৯ | | {a1,a2,b,c,d2,l,g1,e} |

Table 3: Representation of Arabic Numerals

| D Val | Act Sym | Seg Pat | Com Vec | D Val | Act Sym | Seg Pat | Com Vec |
|---|---|---|---|---|---|---|---|
| 0 | ٠ | | { a1,g1,f,I } | 1 | ١ | | { f,e } |
| 2 | ٢ | | { i,g1,f,e } | 3 | ٣ | | { f,e,g1,g2,b,i} |
| 4 | ٤ | | { a1,h,m,d1 } | 5 | ٥ | | { f,e,d1,d2,k,h } |
| 6 | ٦ | | { g2,b,c,i} | 7 | ٧ | | { h,k,c,b } |
| 8 | ٨ | | {m,j,b,c } | 9 | ٩ | | { g1,f,a1,l,l } |

**Table 4: Representation of Greek Numerals**

| D Val | Act Sym | Seg Pat | Com Vec | D Val | Act Sym | Seg Pat | Com Vec |
|---|---|---|---|---|---|---|---|
| 1 | α | | {g1,e,d1,l,g2,k} | 2 | β | | { i,l,a2,j,g2,c,k } |
| 3 | γ | | { h,l,d1,m,j } | 4 | δ | | { h,l,d1,e,g1 } |
| 5 | ε | | { j,g2,k} | 6 | ϛ | | {a1,a2,f,g1,l,d1 } |
| 7 | ζ | | { h,i,a2,j,l,d2 } | 8 | η | | { a1,i,a2,b,c } |
| 9 | θ | | {a2,b,c,d2,l,i,j} | 10 | ι | | { d1,l,d2 } |
| 20 | κ | | { i,l,g2,k } | 30 | λ | | { h,k,m } |
| 40 | μ | | { f,e,g1,i,g2 } | 50 | ν | | { e,m} |
| 60 | ξ | | {h,j,a2,i,g2,l,d2} | 70 | ο | | {a1,a2,b,c,g2,g1,f} |
| 80 | π | | { g1,g2,m,k } | 90 | ϙ | | {a1,a2,b,c,g2,g1,f ,l} |
| 100 | ρ | | {e,f,a1,i,g1 } | 200 | σ | | { j,l,d2,c,g2 } |
| 300 | τ | | { h,g2,b,j,l,d2 } | 400 | υ | | { l,d2,c } |
| 500 | φ | | {a1,a2,b,c,g2,g1,f,l,i} | 600 | χ | | { g1,m,k,j } |
| 700 | ψ | | { f,g1,I,g2,b,l } | 800 | ω | | { f,g1,I,g2,b } |
| 900 | ϡ | | { n, a2,j ,b, c } | | | | |

**Table 5: Representation of Roman Numerals**

| D Val | Act Sym | Seg Pat | Com Vec | D Val | Act Sym | Seg Pat | Com Vec |
|---|---|---|---|---|---|---|---|
| 1 | I | | {a1,a2,d1,d2,l,i} | 5 | V | | {f,e,m,j} |
| 10 | X | | {h,k,m,j} | 50 | L | | {f,e,d1,d2} |
| 100 | C | | {a1,a2,f,e,d1,d2} | 500 | O | | {a1,a2,f,e,d1,d2,b,c} |
| 1000 | M | | {e,f,h,j,b,c} | | | | |

**Table 6: Representation of Chinese Numerals**

| D Val | Act Sym | Seg Pat | Com Vec | D Val | Act Sym | Seg Pat | Com Vec |
|---|---|---|---|---|---|---|---|
| 0 | 零 | | {n,a1,a2,i,m,k,f,b} | 1 | 一 | | {g1,g2} |
| 2 | 二 | | {a1,a2,d1,d2} | 3 | 三 | | {a1,a2,g1,g2,d1,d2} |
| 4 | 四 | | {a1,a2,b,c,d2,d1,e,f,g1,g2,i} | 5 | 五 | | {a1,a2,i,l,d1,d2,g1,g2,c} |
| 6 | 六 | | {g1,g2,m,k,h} | 7 | 七 | | {i,l,d2,g1,g2} |
| 8 | 八 | | {i,m,k,j} | 9 | 九 | | {g1,g2,c,m,i} |
| 10 | 十 | | {g1,g1,l,i} | 100 | 百 | | {a1,a2,b,c,d2,d1,f,e,g1,g2,n} |
| 1000 | 千 | | {a1,a2,i,l,g1,g2} | 10000 | 万 | | {a1,a2,i,g2,c,m} |
| 100000000 | 亿 | | {e,f,a1,a2,j,l,d2,n} | | | | |

## Table 7: Representation of Devanagari Numerals

| D Val | Act Sym | Seg Pat | Com Vec | D Val | Act Sym | Seg Pat | Com Vec |
|---|---|---|---|---|---|---|---|
| 0 | ० | | { a1,g1,f,i } | 1 | १ | | {a1,g1,f,I,m,d1,p} |
| 2 | २ | | {a1,i,l,d1,p} | 3 | ३ | | {a1,i,l,d1,p,g1} |
| 4 | ४ | | { h,l,d1,m,j } | 5 | ५ | | {f,e,d1,l,p} |
| 6 | ६ | | {a1,f,e,d1,g1,p} | 7 | ७ | | {a2,b,c,d2,d1,e,f,i,g2 } |
| 8 | ८ | | {j,m,d1,d2} | 9 | ९ | | {a2,b,g2,i,k,d2} |

## Table 8: Representation of Tamil Numerals

| D Val | Act Sym | Seg Pat | Com Vec | D Val | Act Sym | Seg Pat | Com Vec |
|---|---|---|---|---|---|---|---|
| 1 | ௧ | | {a1,a2,f,e,g1,g2,i,c,p,m} | 2 | ௨ | | {a1,f,g1,i,e,d1,d2} |
| 3 | ௩ | | {e,f,a1,a2,i,g2,c,d2,p} | 4 | ௪ | | {n,f,g1,i,g2,b,l,e,d1} |
| 5 | ௫ | | {a1,a2,g2,f,e,d1,d2,I,l,c} | 6 | ௬ | | {a1,a2,f,e,g1,g2,i,c,m,l} |
| 7 | ௭ | | {e,f,a1,a2,b,c,g1,l,d1} | 8 | ௮ | | {a1,i,l,d1,e,g1,g2,c,b,h} |
| 9 | ௯ | | {e,f,a1,a2,d1,i,l,g1,g2,c,k} | 10 | ௰ | | {f,e,d1,d2,l,i,c,b} |
| 100 | ௱ | | {e,f,a1,a2,b,c,I,l} | 1000 | ௲ | | {e,f,a1,a2,I,g2,c,d2,p,g1,p} |

## Table 9: Representation of Hebrew Numerals

| D Val | Act Sym | Seg Pat | Com Vec | D Val | Act Sym | Seg Pat | Com Vec |
|---|---|---|---|---|---|---|---|
| 1 | א | | { b,e,g1,g2,i,l } | 2 | ב | | {a1,a2,b,c,d1,d2,p } |
| 3 | ג | | {a1,i,l,m,p} | 4 | ד | | {a1,a2,b,c} |
| 5 | ה | | { a2,b,c,l } | 6 | ו | | { b,c } |
| 7 | ז | | {a2,i,l} | 8 | ח | | {a2,b,c,i,l} |
| 9 | ט | | {a2,b,c,d1,d2,e,f} | 10 | י | | {b} |
| 20 | כ | | {a1,a2,b,c,d1,d2} | 30 | ל | | {f,g1,m} |
| 40 | מ | | {a1,a2,b,c,d2,e,f} | 50 | נ | | {a2,b,c,d2} |
| 60 | ס | | {a1,a2,b,c,d1,d2,e,f} | 70 | ע | | {b,c,h,k,d2} |
| 80 | פ | | {a1,a2,b,c,d1,d2,f,g1} | 90 | ץ | | {h,j,l} |
| 100 | ק | | {a1,a2,j,e,f} | 200 | ר | | {a2,b,c} |
| 300 | ש | | {b,c,d1,d2,e,f,g1,i} | 400 | ת | | {a1,a2,b,c,d1,i,l} |

## Table 10: Representation of Gujrati Numerals

| D Val | Act Sym | Seg Pat | Com Vec | D Val | Act Sym | Seg Pat | Com Vec |
|---|---|---|---|---|---|---|---|
| 0 | ૦ | | {a1,a2,b,c,d1,d2,e,f} | 1 | ૧ | | {a1,f,g1,i,l,p} |

| D Val | Act Sym | Seg Pat | Com Vec | D Val | Act Sym | Seg Pat | Com Vec |
|---|---|---|---|---|---|---|---|
| 2 | २ | | {a2,b,g2,k} | 3 | ३ | | {a1,a2,b,c,d1,d2,g2} |
| 4 | ४ | | {d1,h,j,l,m} | 5 | ५ | | {a1,b,c,g2,h} |
| 6 | ६ | | {a1,a2,d1,e,f,g1,g2,p} | 7 | ७ | | {a2,b,c,d1,d2,e,f,i} |
| 8 | ८ | | {d1,d2,j,m} | 9 | ९ | | { d1,d2,e,f,j,m } |

Table 11: Representation of Tibeten Numerals

| D Val | Act Sym | Seg Pat | Com Vec | D Val | Act Sym | Seg Pat | Com Vec |
|---|---|---|---|---|---|---|---|
| 0 | O | | { a1,a2,b,c,d1,d2,e,f } | 1 | ๑ | | {a1,a2,f,j,m} |
| 2 | ๒ | | {a1,a2,b,g2,m} | 3 | ๓ | | {a1,a2,b,c,d1,d2,g2,p} |
| 4 | ๔ | | {a1,d1,f,e,g2,i,p} | 5 | ๕ | | {d1,d2,e,f,i,l} |
| 6 | ๖ | | {a2,c,d1,d2,e,f,i} | 7 | ๗ | | {a1,a2,b,c,d2,e,g2,l} |
| 8 | ๘ | | {j,m } | 9 | ๙ | | {a1,a2,b,d2,e,f,g2,l } |

Table 12: Representation of Telegu Numerals

| D Val | Act Sym | Seg Pat | Com Vec | D Val | Act | Seg Pat | Com Vec |
|---|---|---|---|---|---|---|---|
| 0 | ౦ | | {a1,a2,b,c,d1,d2,e,f } | 1 | ౧ | | {n,h,j} |
| 2 | ౨ | | {a2,b,c,d1,d2,g2,i} | 3 | ౩ | | {a2,b,c,d2,g2} |
| 4 | ౪ | | {b,d1d2,f,g1,g2,k,m} | 5 | ౫ | | {a2,c,g2,h,i,j,km} |
| 6 | ౬ | | {a1,d1,d2,e,f,g1} | 7 | ౭ | | {a1,a2,b,d1,d2,e,g2,g2} |
| 8 | ౮ | | {a2,d1,e,f,i,l} | 9 | ౯ | | {a1,a2,d1,e,c,g1} |

**Segment Patterns Different Mathematical Symbols**

The patterns of different mathematical symbols are shown in table. In the following table 13 "Act Sym" stands for Actual Symbol, "Seg Pat" stands for Segment Pattern, "Com Vec" stand for Combination Vector.

Table 13: Representation of Mathematical Symbols

| Act Sym | Seg Pat | Com Vec | Act Sym | Seg Pat | Com Vec | Act Sym | Seg Pat | Com Vec |
|---|---|---|---|---|---|---|---|---|
| Plus | | { i,1,g1,g2} | Minus | | { g1 } | Multiply | | { h,j,l,m } |
| Divide | | { j,m } | Plus-minus | | {il,g1,g2,d1,d2 } | Percentage | | {a1,f,g1,i,g2,c,l,d2,j,m} |
| Left 1st Brace | | { m, h } | Right 1st Brace | | { j, k } | Left 2nd Brace | | { a1,i,g2,l,d1 } |
| Right 2nd Brace | | {a2,i,g1,l,d2 } | Left 3rd Brace | | {a2,d2,i,l} | Right 3rd Brace | | { a2,i,l,d2 } |
| Dot | | { g1,l,d1,e} | Comma | | { g1,m } | Degree | | {a2,b,g2,i } |
| Radian | | {a2,i,g2} | Prime | | { j } | By | | { l,d1 } |
| Greater Equal | | { h,g1,d1 } | Greater | | { h,g1 } | Smaller Equal | | { j,g2,d2 } |

| Act Sym | Seg Pat | Com Vec | Act Sym | Seg Pat | Com Vec | Act Sym | Seg Pat | Com Vec |
|---|---|---|---|---|---|---|---|---|
| Smaller | | { j,g2 } | Equal | | { g1,d1 } | Equality | | {a1,g1,d1} |
| Not Equal | | {j,m,d1,d2,g1,g2} | Tilde | | {f,h,g2,b} | Union | | {i,l,d2,c,b} |
| Intersection | | { i,l,a2,c,b} | Subset | | {d1,d2,e,g1,g2} | Super Set | | {d1,d2,c,g1,g2} |
| Element of | | {a1,a2,d1,e,f,g1,g2} | Not An Element | | {a1,a2,f,g1,g2,e,d1,d2,m,j} | Right angle | | {g1,g2,m,j} |
| Angle | | {g1,g2,i,l} | Left Ceil | | { a2,l,i } | Right Ceil | | { a1,i,l} |
| Left Floor | | { h,g,f } | Right Floor | | {d2,i,l} | Pi | | {g1,g2,m,k } |
| Infinity | | {g1,g2,d1,d2,c,e} | Sum | | {a1,a2,h,m,d1,d2} | Power | | { l,m } |
| Factorial | | { f,e,d1,d2 } | Sigma | | { a1,a2,b,c} | Delta | | {m,d1,d2,k} |
| Inverse Delta | | {a1,a2,h,j} | Differentiation | | {i,l,d1,e,g1} | Integration | | {a2,i,l,d1,g1,g2} |
| For All | | {f,e,m,j,g1} | Parallel | | {i,l,b,c} | Square Root | | { e,m,j } |
| Or | | {h,k,b,c} | And | | {m,j,b,c} | | | |

## APPLICATIONS

As the 18-segment display is based on 16-segment display it can represent Latin alphabets too. In Greek language the alphabet and numerals are same, that's why this display can be used to display Latin and Greek word and also can be used to display numeric expressions.

## CONCLUSIONS

In this paper we have proposed 18-segment display for representing numerals of twelve different language numeric symbols which is used than 50 languages and can also display mathematical expressions. As this display architecture supports multiple language numerals together, it can be considered as the simplest universal display. In future we will try to improve its usability by displaying numerals of other languages.